\documentclass[aps,prl,twocolumn,superscriptaddress,floatfix]{revtex4-1}
\usepackage{amsmath,amssymb,graphicx,graphics,color}
\usepackage{dcolumn}
\usepackage{bm}
\usepackage{psfrag}

\begin{document}
\title{Universal logarithmic temperature dependence of magnetic susceptibility of one-dimensional electrons at critical values of magnetic field}   
\author{T. Vekua}
\affiliation{Institut f\"ur Theoretische Physik, Leibniz Universit\"at 
Hannover, 30167~Hannover, Germany}

\begin{abstract}
We study the leading low temperature dependence of magnetic susceptibility of one-dimensional electrons with fixed total number of particles at the magnetic fields equal to zero temperature critical values where magnetic field induces commensurate-incommensurate quantum phase transitions.
For free and repulsively interacting electrons there is only one such critical field corresponding to the transition to the fully polarized state. For attractively interacting electrons besides saturation field there is another critical field equal to the spin gap where zero temperature magnetization sets in. For all cases, except of the lattice models at half filling, the magnetic susceptibility at critical values of magnetic field has a universal logarithmic temperature dependence, $\chi(T)=\chi(0)(1+2/\ln{T}+\cdots)$ for $T\to 0$. 
\end{abstract}

\maketitle

 The simplest quantum phase transition\cite{Sachdev0} is a transition between the particle vacuum and a finite density state driven by changing the chemical potential. In one-dimensional (1D) systems such transition, if density does not experience macroscopic jump across the transition, is termed commensurate- incommensurate (C-IC) \cite{Japaridze,Pokrovski} phase transition and is of second order. Density increases as a square root of change of chemical potential in the vicinity of the transition. Simplicity of C-IC transition is connected with the fact that it is captured by free fermions. Dominance of kinetic energy over potential energy at vanishingly low particle density in short-range interacting systems makes C-IC transition universal (potential energy must fall faster than $1/x^2$, hence unscreaned Couplomb interactions are excluded). C-IC transition obeys a zero scale-factor universality \cite{Sachdev} and is characterized by a dynamical critical exponent $z=2$. At the C-IC phase transition point the chemical potential touches the bottom of the empty particle band and since particles disperse quadratically at the bottom of the band, in 1D there is a van-Hove singularity in the density of states. In particular for the specific heat it implies square root temperature dependence $C(T)\sim \sqrt{T}$ for $T\to 0$ and the density susceptibility (derivatve of particle density with respect to the chemical potential) shows square root singularity $\chi(T)\sim 1/\sqrt{T}$. For single component systems, like spin polarized electrons, or repulsively interacting single species Bose gas or majority of spin chains C-IC transition is defined for grand canonical systems, as it involves change of particle density.

C-IC transitions are frequent as well in multicomponent systems, with classic example provided by spinful electrons in external magnetic field. Magnetization process at zero temperature for attractive electrons is depicted in Fig. 1. Here external magnetic field induces two C-IC transitions. In particular at point $h=h_{c1}$ the transition is from the zero magnetization state to a state with finite magnetization and at $h=h_{c2}$ system is fully polarized.
When one discusses C-IC phase transition for multicomponent systems, like spinfull electrons or multicomponent repulsive Bose gas, one has to specify whether system is canonical, with fixed total number of particles or grand canonical, connected to external particle reservoir. It was shown that in multicomponent canonical systems the square root dependence of specific heat at C-IC phase transition points is generically modified to a linear one with logarithmically divergent prefactor (with an exception of lattice models at half-filling) \cite{me}, 
\begin{equation}
\label{specificheat}
 C(T)\sim {T\ln^2{T}}
\end{equation}
 in the limit $T\to 0$. 

  System of free electrons at the saturation value of magnetic field, with a constraint that total number of fermions is kept constant explains modification of square root dependence to the one given in Eq. (\ref{specificheat}). Namely, van-Hove singularity due to quadratically dispersing down-spin electrons (assuming that magnetic field favours up-spin electrons) is constrained by the unwillingness to deplete the band of linearly dispersing up-spin electrons. Lagrange multiplier, temperature dependent chemical potential that keeps total number of particle fixed $N_{\uparrow}(T)+N_{\downarrow}(T)=const$, picks up logarithmic dependence at low temperatures \cite{me} that eventually carries to specific heat.
Interestingly van-Hove singularity can be constrained not only due to keeping total particle number fixed (quadratic in fermions terms), but also, as argued due to interactions (quartic in fermions terms) between quadratically and linearly dispersing modes and specific heat again changes from square root to linear temperature dependence with a diverging prefactor \cite{Rosch}.

Natural question is how magnetic susceptibility's $1/\sqrt{T}$ divergence in $T\to 0$ limit, a hallmark behavior of C-IC phase transition for grand canonical systems is modified by the constraint of keeping total number of particles fixed at multicomponent C-IC phase transitions. We are going to unswer this question in our work.

When describing behavior of ground state magnetic susceptibility near the C-IC phase transition point one can not rely on mode-decoupling
approximation a la spin-charge separation \cite{Vekua1}. In particular, magnetic susceptibility of spin gapped system of attractively interactong electrons, at the edge point of magnetization plateau, stays finite instead of diverging when total number of electrons is kept constant, unless special microscopic symmetries are present \cite{Vekua1} (e.g. particle-hole symmetry at half filling for attracting Fermi
Hubbard model). Similar behavior has also been observed for a number of
exactly solvable models like the two component Fermi Hubbard model both on lattice\cite{Woynarovich86,FrahmVekua} as well as in continuum\cite{Orso}, or spin $S$ generalization of the integrable $t-J$ chain doped with $S-1/2$ carriers\cite{Frahm}, antiferromagnetically interacting spin-$1$ Bose gas\cite{Guan}, etc.

\begin{figure}
\begin{center}
\includegraphics[width=7.5cm]{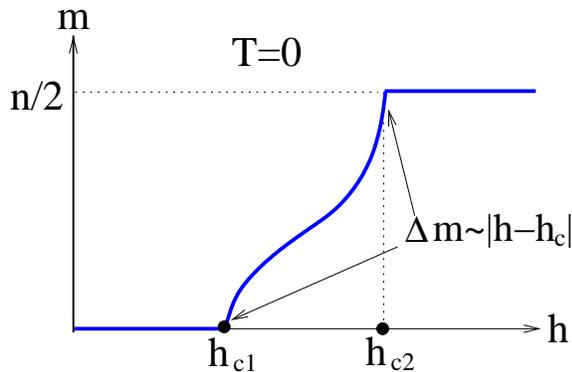}
\end{center}
\caption{Ground state magnetization curve of attractive fermions for canonical system. At the edges of magnetization plateau susceptibility is finite and magnetization increases linearly with field, $\Delta m =\chi(0)|h-h_{c}|$. For repulsively interacting and free fermions $m=0$ plateau is not present ($h_{c1}$ point does not exist). At $h_{c2}$ for lattice electrons $\chi(0) \sim \frac{1}{4\pi \sin( \pi n)}$ (proportionality constant is interaction dependent), where $n$ is lattice filling. For $n=1$, at half filling $\chi(0)$ diverges at saturation magnetic field $h_{2c}$ (and also at $h_{c1}$ for attractive interactions) as $\chi(T)\sim 1/\sqrt{T}$.}
\label{fig:1}
\end{figure}

 To our knowledge no previous studies on finite temperature behavior of susceptibilities at C-IC phase transition points have been reported for multicomponent canonical systems.
Neither are we aware of the results for the leading temperature dependence of magnetic susceptibility for the textbook case of canonical 1D free electrons (in lattice or in continuum) at C-IC point at the saturation magnetic field value. We remind that saturation field value applies to $h_{c2}$ on Fig. 1 at $T=0$, since at non-zero temperature saturation field shifts to infinity. 

For concrete calculations we will consider a system of two component
attractive fermions and place it in an external magnetic field equal to the
value of the zero temperature spin gap, denoted by $h=h_{c1}$ in Fig. 1.

\begin{eqnarray}
H=&-&t\sum_{j,\sigma=\{\uparrow, \downarrow\}}( c^{\dagger}_{\sigma,j}c_{\sigma,j+1}+ h.c.)+4u\sum_jn_{\uparrow,j}n_{\downarrow,j} \nonumber\\
&-&{h}\sum_j \frac{n_{\uparrow,j}-n_{\downarrow,j}}{2}.
\end{eqnarray}
We will fix $t=1$ throughout the paper and measure interaction $u$ magnetic field $h$ and temperature $T$ in units of $t$. We will show that due to keeping total number of particle fixed the magnetic susceptibility will approach its zero temperature finite value with a universal logarithmically divergent slope at low temperatures. Since we will use only basic properties of low energy dispersion of the relevant modes, our main claim on magnetic susceptibility at C-IC phase transition
point will hold true for generic situation (and in particular at saturation magnetic field value $h=h_{c2}$ in Fig. 1), wherever quadratically dispersing soft mode at the critical point couples to linearly dispersing excitations by contraint (charge mode). This statement will apply to all situations where magnetic susceptibility stays finite at the edge points of the magnetization plateaus in zero temperature limit. 

We will follow closely and extend calculations presented in \cite{me}. We will repeat some of the steps shown in for \cite{me} for clarity and the sake of completeness.
To simplify things drastically we will consider the dilute strong coupling limit of the
system of attractively interacting two component fermions in continuum. From the Bethe ansatz solution it is well known that case for $h\le h_{c1}$ the ground state is made of bound pairs and the low temperature thermodynamic properties can be modeled by the mixture of noninteracting pairs and thermally
created uncompensated up-spin particles \cite{Woynarovich86}. Down-spin particles
will also be created thermally, but since they have spectral gap (which is
large in strong coupling) their density will be exponentially suppressed at
low temperatures and thus they will be ignored in the following. For
integrable Fermi Hubbard model (both on lattice\cite{Essler} and in continuum)
the thermodynamic Bethe ansatz method can be applied \cite{Takahashibook}, and in particular recently it was shown that in dilute limit this method simplifies considerably\cite{Zhao}. However, for finding leading temperature dependence of magnetic susceptibility at the magnetic field induced C-IC phase transition points we need not use integrability so that our reasoning will stay both simple and general.

Hamiltonian describing the low energy properties of the two component attractive fermions in strong coupling limit $n\to 0, u<0$ near the C-IC critical point reads:
\begin{equation}
\label{PhenHam}
 H\!=\!\!\!\sum_k \!(E_p(k)-E_F)a^{\dagger}_{k} a_{k} +\!\!\sum_k\! (E_{\uparrow}(k)\!-\frac{h-h_{c1}}{2})c^{\dagger}_{k} c_{k},
\end{equation}
where $a^{\dagger}(c^{\dagger})$ and $a(c)$ are pair (ucnompensated, excess up-spin electrons) creation and annihilation operators. Note, magnetic field does not couple to electron pairs, since pairs are spin neutral.
For calculating thermodynamic properties it does not matter whether pairs are modeled by hard core bosons or fermions\cite{Takahashibook}.

\begin{figure}
\begin{center}
\includegraphics[width=5.0cm]{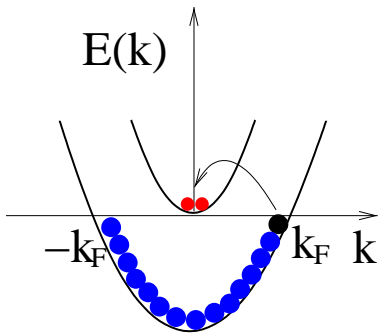}
\end{center}
\caption{Illustration of the constraint of keeping total number of electrons fixed: creation of 2 uncompensated up-spin electrons involves leaving of a single hole in the Fermi sea of electron pairs.}
\label{fig:2}
\end{figure}
At the edge of the zero magnetization plateau ($h=h_{c1}$ in Fig. 1) the low energy dispersion of excess up-spin particles is quadratic,
\begin{equation}
 E_{\uparrow}(k)= k^2.
\end{equation}
We have fixed units $\hbar=1$ and mass of electrons $M=1/2$.
 The low energy dispersion of pairs though is linear in momentum: 
\begin{equation} 
E_p(k)-E_F=\frac{k^2}{2}-E_F\simeq v_p(|k|-k_F)\,\,   \mathrm{for}\,\, |k|\to k_F,
\end{equation}
where in the units adopted by us velocity of the pairs at Fermi energy is $v_p=k_F$, with $k_F$ being a Fermi wave vector of electron pairs (see Fig. 2) that is related to the linear density of pairs $n_p$ by the standard 1D relation $k_F=n_p \pi$ (we denote by $n_p$ the density of pairs at zero temperature: $n_p=n_p(T=0)$). Mass of the pair is double mass of single fermion $m_p=2M=1$. 
At zero temperature at critical field $h=h_{c1}$ the ground state is occupied solely by pairs, and density of uncompensated up-spin electrons is zero. 
Since each pair destroyed by thermal fluctuation or by magnetic field produces two up-spin electrons, as depicted on Fig.(\ref{fig:1}), the constraint that fixes the total number of particles reads:
\begin{equation}
\label{con}
n_{\uparrow}(T)+2n_p(T)=const.
\end{equation}

To implement the constraint we introduce a Lagrange multiplier, that is a temperature dependent chemical potential. It is a solution of the following equation,
\begin{equation}
\int\!\!\! \frac{dk/2\pi}{ e^{\frac{ E_{\uparrow}(k)-\varepsilon-\mu(T) }{T}} +1}+\int\!\!\!\frac{dk/\pi}{ e^{\frac{ E_p(k)-E_F-2\mu(T) }{T}} +1}=const.
\end{equation}
Note, $2\varepsilon=h-h_{c1} \to 0$ is an infinitezimal deviation of magnetic field from the critical value and we have set $k_B=1$. For $T\to 0$ the leading temperature dependence of Lagrange multiplier can be obtained by solving the following equation,
\begin{equation}
\label{Lambert}
\mu_{\varepsilon}(T)= -\alpha \sqrt{T}e^{[{\varepsilon+\mu_{\varepsilon}(T)]}/{T}}.
\end{equation}
where we have introduced notation $\alpha=\sqrt{\pi} v_p/8.$

Solution of Eq. (\ref{Lambert}) is given in terms of product logarithm function $\mathrm{P}(z)$, also known as Lambert $\mathrm{W}$ function,
\begin{equation}
\mu_{\varepsilon}(T)/T=-\mathrm{P}\Big(\frac{ \alpha e^{{\varepsilon}/{T}} }{\sqrt{T}}\Big).
\end{equation}
As already mentioned, at critical field $\varepsilon=0$, the Lagrange multiplier chemical potential picks up logarithmic dependences on temperature \cite{me} for $T\to 0$,
$$
 \mu_0(T)\simeq -T(\ln{\frac{\alpha}{\sqrt{T}}} -\ln \ln{\frac{\alpha}{\sqrt{T}}}+ \frac{\ln \ln{\frac{\alpha}{\sqrt{T}}}}{\ln{\frac{\alpha}{\sqrt{T}}}})+\cdots
$$
where dots here and after indicate sub-leading in temperature terms. Logarithms appearing in Lagrange multiplier carry to thermodynamic quantites, i.e. specific heat in Eq. (1). One can anticipate appearance of $\ln {T}$ contributions in magnetic susceptibility as well. In the following we explicitely confirm the above expectation. 

Magnetic susceptibility expressed in terms of uncompensated up-spin electron density is 
\begin{equation}
\chi(T)= \frac{ \partial n_{\uparrow}(T)}{2\partial h} .
\end{equation}
We remind that $ n_{\uparrow}(T) $ is the density of uncompensated up-spin electrons, whereas number of up-spin electrons is $N_{\uparrow}(T)=L(n_{\uparrow}(T)+n_p(T))$, where $L$ is system size, since each pair contains one up-spin (as well as one down-spin) electron.

We are interested in particular in the behavior of magnetic susceptibility at the critical magnetic field values,
\begin{equation}
\chi_c(T)=\frac{\partial n_{\uparrow}(T)}{4\partial \varepsilon}|_{\varepsilon=0}.
\end{equation} 

Density of uncompensated up-spin electrons at $h=h_{c1}+2\varepsilon$ is,
\begin{equation}
 n_{\uparrow}(T)=-\sqrt{T/4\pi}\mathrm{Li}_{\frac{1}{2}}(-e^{[\varepsilon+\mu_{\varepsilon}(T)]/T}),
\end{equation}
where $\mathrm{Li}_{\frac{1}{2}}(z)$ is a polylogarithm function originating from a standard Fermi-Dirac integral.
After some lengthy but straightforward simplifications for canonical electrons (with fixed total number of particles) in the limit of $T\to 0$ for the magnetic susceptibility's temperature dependence at $h=h_{c1}$ we obtain,
\begin{eqnarray}
\label{mainr}
\frac{\chi_c(T)}{\chi_c(0)}&=& 1- \frac{1}{1+\mathrm{P}({\alpha}/{\sqrt{T}})}+\cdots \\
&\simeq &   1+  \frac{2}{\ln({T/\alpha^2})} \left(1+\frac{2(1-\ln\ln{\alpha /\sqrt{T}})} {\ln (T/ \alpha^2)}\right)  +\cdots , \nonumber
\end{eqnarray}

where $\chi_c(0)=1/ \pi v_P$ is an exact value of zero temperature  magnetic susceptibility  of attractive fermions at $h_{c1}$ in the strong coupling limit \cite{Woynarovich86,Vekua1,Orso} known from Bethe ansatz solution. Subindex $c$ of susceptibility indicates that magnetic field strength equals to critical value corresponding to zero temperature C-IC phase transition. Note that zero temperature susceptibility is a discontinuous function of magnetic field, experiencing jumps at $h_{c1}$ from zero to finite value with increasing the field and at $h_{c2}$ from finite value to zero as one can deduce from Fig. 1. By $\chi_c(0)$ we understand right derivative of magnetization with respect to magnetic field for $h_{c1}$ point and the left derivative for $h_{c2}$ point. However, for any $T>0$ magnetic susceptibility is a continuous function of magnetic field hence finite temperature derivatives of magnetization at $h_{c1}$ and $h_{c2}$ are uniquely defined.

Eq. (\ref{mainr}) is our main result. Namely for canonical systems magentic susceptibility approaches zero temperature value with a logarithmic in temperature slope as depiced in Fig. 3,
\begin{equation}
\frac{\partial \chi_c(T)}{\partial T}\sim -\frac{1}{T \ln^2{T}}+\cdots \sim -{C^{-1}(T)}.
\end{equation}

Similar infinite slope of magnetic susceptibility at $T=0$  was obtained for XXZ antiferromagnetic spin-$\frac{1}{2}$ chain \cite{Eggert} at zero magnetic field in the vicinity of SU(2) antiferromagnetic point in the gapless region. However the slope of magnetic susceptibility in XXZ chain is positive and its origin is interpreted in field theory formulation due to the presence of irrelevant operators in effective description of spin-$\frac{1}{2}$ chain. In our case the origin of logarithmic slope of magnetic susceptibility is much simpler, as it is related to the constrained van-Hove singularity in the density of states of quadratically dispersing spin mode due to linearly dispersing charge mode (constraint arises
 when keeping total number of particles fixed) and hence is generic. Magnetic susceptibility of free 1D electrons at saturation magnetic field displays same behavior when total number of particles is fixed except for the half-filled bands. 

\begin{figure}
\begin{center}
\includegraphics[width=8.0cm]{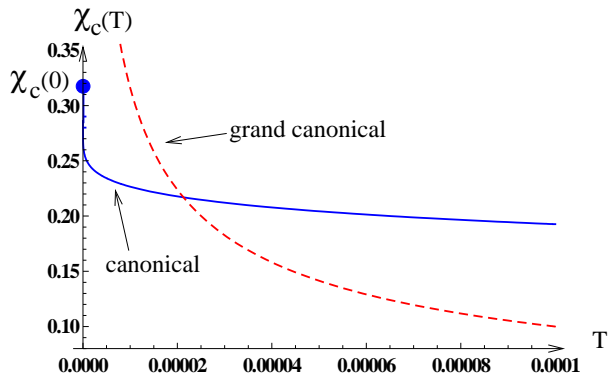}
\end{center}
\caption{Qualitative comparison of magnetic susceptibility of canonical and grand canonical one-dimensional electrons at low temperatures at critical magnetic field inducing C-IC phase transition. Continuous line: result for canonical systems for $v_p=1$. Dashed line: inverse squre root dependence for grand canonical systems $10^{-3}/\sqrt{T}$.}
\label{fig:4}
\end{figure}

To summarize within the free fermion framework we have obtained the leading temperature dependence of magnetic susceptibility of two-component canonical electrons at magnetic fields corresponding to zero temperature critical values, $\chi_c(T)=\chi_c(0)(1+2/\ln{T}+\cdots)$.
Only arguments in our analyses we used was existence of two types of dispersion, linear and quadratic at low energies and a constraint between the two. Hence our result for magnetic susceptibility is very robust and generic. Together with the previously established result of specific heat \cite{me} $C(T)\sim T \ln^2{T}$, the obtained low temperature behavior of magnetic susceptibility concludes basic finite temperature characterization of C-IC quantum critical point for one-dimensional canonical multi-component systems.

Our findings may be relevant for realistic one-dimensional electron systems, where one can measure experimentally magnetic susceptibility. Square root divergence, which is expected at half filling, should be modified by logarithmic dependence given in Eq. (12) away from half filling, at critical magnetic field strengths i.e. at the saturation magnetic field for repulsively interacting electrons.

\section{Acknowledgments}

I thank H. Frahm for discussions. The author is supported by Center for Quantum Engineering and Space-Time Research (QUEST) and DFG Research Training Group (Graduiertenkolleg) 1729.


\end{document}